\documentclass[11pt,twoside]{article}
\usepackage{asp2014}

\aspvolume{512}
\aspcpryear{2017}
\aspvoltitle{Astronomical Data Analysis Software and Systems XXV}
\aspvolauthor{Nuria P.F. Lorente, Keith Shortridge, and Randall Wayth, eds.}

\resetcounters

\begin{document}

\title{Data Analysis for Precision Spectroscopy: the ESPRESSO Case}
\author{Guido~Cupani$^1$, Valentina~D'Odorico$^1$, Stefano~Cristiani$^1$, Jonay~Gonz\'alez-Hern\'andez$^2$, Christophe~Lovis$^3$, S\'ergio~Sousa$^4$, Paolo~Di~Marcantonio$^1$, and Denis~M\'egevand$^3$
\affil{$^1$INAF--OATs, Via Tiepolo 11, 34143 Trieste, Italy}
\affil{$^2$IAC, V\'ia L\'actea, 38205 La Laguna, Tenerife, Spain}
\affil{$^3$Universit\'e de Gen\`eve, 51 Chemin des Maillettes, 1290 Versoix, Switzerland}
\affil{$^4$Instituto de Astrof\'isica e Ci\^encias do Espa\c{c}o, Universidade do Porto, CAUP, Rua das Estrelas, 4150-762 Porto, Portugal}}

\begin{abstract}
Astronomical Spectroscopy is rapidly evolving into a precision science, with several science cases increasingly relying on long-term instrumental stability and centimeter-per-second accuracy in wavelength calibration. These requirements strongly call for integrated software tools to manage not only the reduction of data, but also the scientific analysis. The ultra-stable, high-resolution echelle spectrograph ESPRESSO, currently under integration for the ESO Very Large Telescope (first light: 2017) is the first instrument of its kind to include a dedicated Data Analysis Software among its deliverables, to process both stellar and quasar spectra. This software will extract physical information from the reduced data on the fly (e.g., stellar radial velocities, or characterisation of the absorption systems along the sightline to quasars) and will allow interaction through a configurable graphical user interface. In this article we present the features of the ESPRESSO Data Analysis Software and its development status at the first complete internal release. A particular attention is devoted to the algorithms developed for quasar spectral analysis (continuum determination and interpretation of the absorption systems). 
\end{abstract}

\section*{ESPRESSO in the age of precision spectroscopy}
A number of observational projects in the field of optical spectroscopy are currently pushing the limits of the current instrumentations, in terms of sensitivity, thermo-mechanical stability, and precision and accuracy of the wavelength calibration. A few of them are:
\begin{itemize}
\item\emph{The search of extrasolar planet with the method of the radial velocity}: this technique, aimed at detecting periodic fluctuations in the radial velocity of stars, can be in principle be used to detect Earth-like planets in the habitable zone, but it requires an accuracy below $10$ cm s$^{-1}$. Current ultra-stable spectrographs like HARPS-N \citep{2012SPIE.8446E..1VC} fall short of this limit by about one order of magnitude;
\item\emph{The detection of a possible variation in the value of dimensionless physical constants} (namely, the fine structure constant $\alpha$ and the proton-to-electron mass ratio $\mu$): such a variation would leave an imprint in the relative position of absorption lines at the same redshift, likely due to the same gas cloud, along the line of sight to distant sources (e.g., quasars). For both constants, different authors claim a variation of some parts per million (e.g.~\citealt{2011PhRvL.107s1101W,2012PhDT........14K,2013MNRAS.435..861R}), but the results are still debated and would benefit from a $10$ cm s$^{-1}$ precision in wavelength calibration.
\item\emph{The direct measurement of the expansion of the universe from a redshift drift of distant sources} (the so-called Sandage test, \citealt{1962ApJ...136..319S}), which would provide a totally independent confirmation to the concordance cosmological model. The velocity variation to be revealed is of the order of $1$ cm s$^{-1}$ yr$^{-1}$.
\end{itemize}
The last science case is beyond the capacities of the current facilities and requires thousands of hours of observations appropriately spread across a time scale of at list two decades with the next generation of extremely large telescopes (e.g., the ESO E-ELT; \citealt{2008MNRAS.386.1192L}). The other two cases can be assessed by exploiting the existing telescopes to their full potential, and provide an optimal starting point to pave the way towards the ELT era.

The new instrument ESPRESSO \citep{2014AN....335....8P} has been purposely designed to cope with these requirements. It is a fiber-fed echelle spectrograph for the coudé combined focus of the ESO VLT, and it will be able to use the light from all four unit telescopes at once. The combination of three nested thermal enclosures and a vacuum vessel will stabilize the temperature and the pressure at the echelle grating with great accuracy ($1$ mK and $1$ $\mu$Pa in a 1-night time scale). High and ultra-high resolution modes ($R=55,000$-$200,000$) will be available across the whole visual band ($380$-$780$ nm). The wavelength calibration will be provided by a laser-frequency comb to achieve a precision of $10$ cm s$^{-1}$ \citep{2012Natur.485..611W, 2012Msngr.149....2L}. The first light is foreseen in 2017. 

\section*{The ESPRESSO DAS in the ESO-Reflex environment}

\articlefigure[width=\textwidth,height=0.55\textheight]{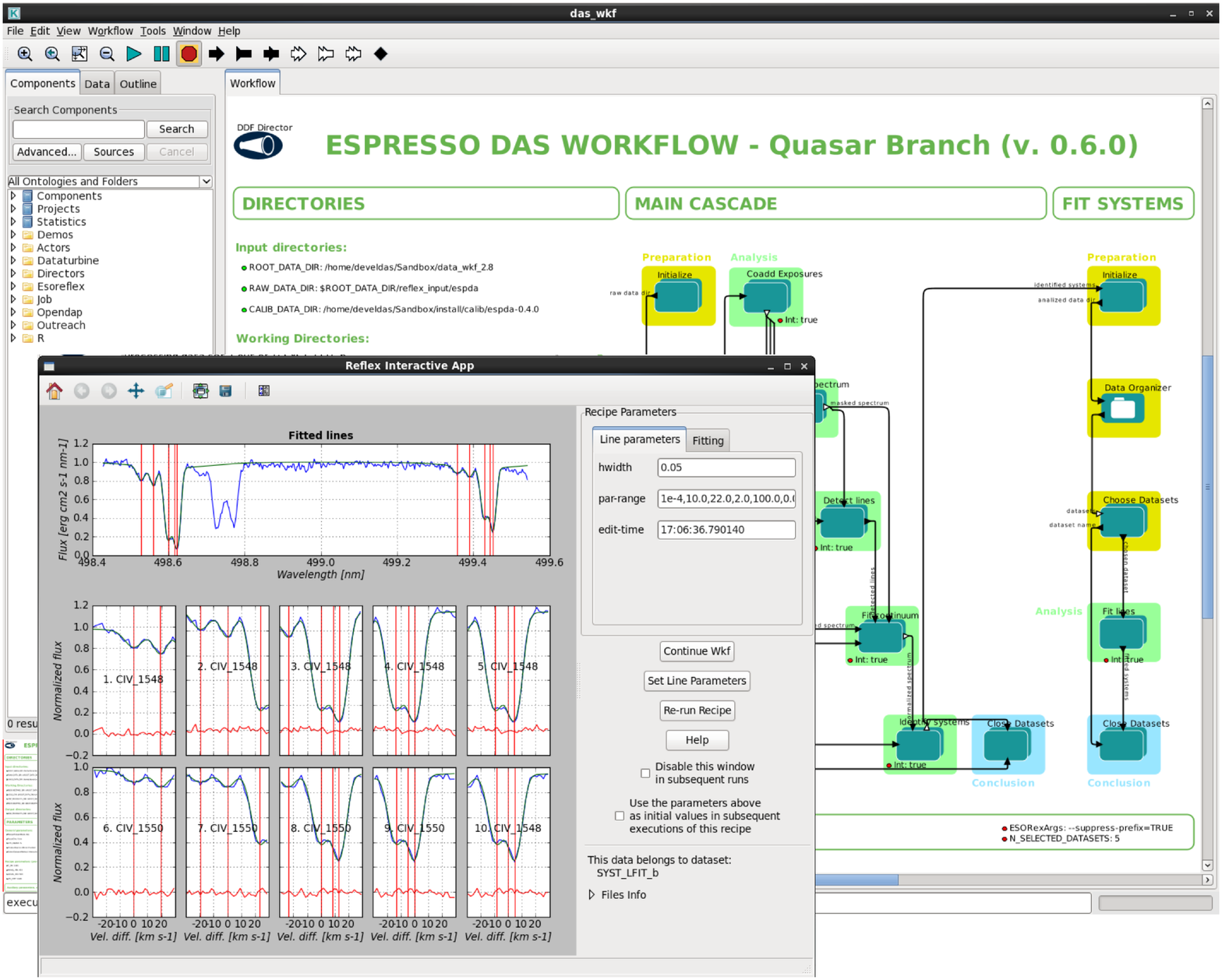}{ESO_reflex}{ESO Reflex workflow for the quasar branch of the DAS. \emph{Background}: main canvas, showing the available operations, properly connected to direct data through the recipes. \emph{Front}: dialog window of the line-fitting recipes. An absorption systems consisting of a C \textsc{iv} doublet $\lambda\lambda 1548$, $1550$ with five identified component has been fitted with Voigt profiles. Constraints have been set to fit each C \textsc{iv} $\lambda 1548$ components  with the same parameters of the corresponding C \textsc{iv} $\lambda 1550$ components. A pop-up editable table (not shown) allows to manually adjust these constraints and to change the guess values of line parameters to be used in the fitting procedure.}

The very specific science cases of ESPRESSO, together with its state-of-the-art technological specifications, call for a dedicated tool for data treatment, encompassing both the removal of the instrumental signature (i.e., the \emph{data reduction}) and the extraction of scientific information from the reduced data (i.e., the \emph{data analysis}). ESO has been mantaining since two decades an advanced Data Flow Systems (DFS) to control all operations on data, from observation preparation to data reduction, with multiple feedback tools all along the chain. All the operations have been progressively covered by \emph{ad-hoc} software tools, maintained and validated by ESO, except for the data analysis part, which is still left to the end users (who typically employ non-automatic, non-validated procedures). With ESPRESSO the situation is going to change.

In addition to the standard Data Reduction Software (DRS), ESPRESSO will be equipped with a Data Analysis Software (DAS) to treat both stellar and quasar spectra. Both software suites are made up of indipendent ANSI-C modules (called \emph{recipes}) based on the ESO Common Pipeline Library (CPL; \citealt{2004SPIE.5493..444M}) to perform separate analysis procedures for both stellar and quasar spectra. The main characteristic of data analysis is that it is not a linear procedures: the various operations are entangled with each other, and the results are to be refined through successive iterations (a picture quite different from the common ``pipeline'' paradigm adopted by the DRS). To address this issue, the DAS takes advantage of ESO Reflex environment \citep{2013A&A...559A..96F}, a graphical user interface developed on the Kepler workflow engine to handle the recipes in an automatic yet controlled way. Fig.~\ref{ESO_reflex} shows some snapshots of the ESO Reflex workflow for the quasar branch of the  DAS. The green boxes are ``actors'' responsible for retrieving the input data, executing the recipes, and save the products. The main operations performed by this workflow are the following:
\begin{itemize}
\item\emph{Co-addition of multiple exposures}, which produces two outputs: a rebinned spectrum, obtained by combining the exposures into a single table with a fixed wavelength grid, and a collated table with the original information (flux, error on flux and calibrated wavelength) from the detector pixels;
\item\emph{Detection of the absorption lines} as outliers in the distribution of flux values within small wavelength chunks;
\item\emph{Fitting of the continuum component in emission}, which is performed by an incremental removal of the lines (modeled with Voigt profiles); in the so-called ``Lyman-$\alpha$ forest'', a corrective term computed from the column density distribution of the removed lines is applied to the flux to take into account also the absorption not associated with detectable lines;
\item\emph{Identification of the absorption systems} through the evaluation of possible redshift coincidences among lines (with an algorithm derived from \citealt{1975ApJ...198...13A});
\item\emph{Voigt-profile fitting of the identified absorption systems}, which refines the information about the redshift, the column density, and the thermal broadening of single lines, while enforcing the required constraints between them.
\end{itemize}
ESO Reflex is equipped with user-friendly interactive tools to display the results and adjust the parameters through mouse and keyboard actions. An example of interactive window (a zoomable, clickable array of plots, with a panel to edit parameters and repeat the execution) is shown in the front window of Fig.~\ref{ESO_reflex}. 

The most important novelty of the code is the fact that non-rebinned reduced data are propagated throughout the analysis process alongside the standard rebinned spectrum obtained from coaddition. Tests on simulated absorption lines show that the results of line fitting performed on an ensemble of rebinned spectra do not follow a chi-squared distribution (Cupani et al., in prep). In fact, errors in the rebinned spectra are correlated across pixels and the degrees of freedom are ill-defined due to degeneration among the line parameters. The problem is avoided when using non-rebinned spectra; this is the only safe condition in which a chi-squared best-fitting test can be applied, and it will be first made available for quasar spectral analysis by the ESPRESSO DAS.

\bibliography{O2-2} 

\end{document}